\documentclass[a4paper,12pt,titlepage]{article}
\usepackage{a4wide}
\usepackage{amsfonts}
\usepackage{amsmath}
\usepackage{mathrsfs}
\usepackage{txfonts}
\def\C{\mathbb{C}}
\def\A{\mathscr{A}}

\def\HF{\mathbb{F}}
\def\CP{\mathbb{CP}}
\def\N{\mathscr{N}}

\def\F{\mathscr{F}}
\def\A{\mathscr{A}}
\def\G{\mathscr{G}}
\def\M{\mathscr{M}}
\def\S{\mathfrak{S}}
\def\X{\mathbb{X}}
\def\etal{\textit{et.\ \!al.}\ }
\def\ala{\textit{\`a la}\ }
\DeclareMathOperator{\Det}{Det}

\DeclareMathOperator{\tr}{tr}
\DeclareMathOperator{\Ind}{Ind}
\DeclareMathOperator{\diag}{diag}

\DeclareMathOperator{\Ker}{Ker}
\def\five-dimensional{5D}
\def\four-dimensional{4D}
\def\slash#1{\ooalign{{\text{$#1$}}\crcr \hss\big/\hss}}
\def\eqref#1{(\ref{#1})}
\begin{document}
\baselineskip.99\baselineskip
\begin{titlepage}
\begin{flushright}
UT-04-01\\
hep-th/0401184
\end{flushright}
\vfil
\begin{center}
\LARGE Five-dimensional Chern--Simons terms\\
and Nekrasov's instanton counting\\
\bigskip
\large {Yuji Tachikawa}\\
\bigskip
\textit {Department of Physics, Faculty of Science}\\
\textit {University of Tokyo, 113-0033 Tokyo JAPAN}
\end{center}
\bigskip\bigskip\bigskip\bigskip\bigskip\bigskip
\centerline{\large\textbf{abstract}}
\bigskip
We extend the graviphoton-corrected prepotential  of \five-dimensional\ 
pure $U(N)$  super Yang-Mills, which was originally proposed by Nekrasov,
by incorporating the effect of the \five-dimensional\ Chern-Simons term.
This extension allows us to reproduce by a gauge theory calculation
the partition functions of corresponding topological A-model on 
local toric Calabi-Yau manifolds $X^m_N$ for all $m=0,1,\ldots,N$.
The original proposal corresponds to the case  $m=0$.
\end{titlepage}

\section{Introduction}
The determination of low-energy prepotential of four-dimensional (\four-dimensional)
$\N=2$ super Yang-Mills theory initiated by Seiberg and Witten
marked a  significant step toward our understanding of non-perturbative dynamics
of  gauge theory.
The results were  originally derived  by exploiting holomorphy 
and electromagnetic duality.
Recently,
several authors\cite{FlumeLocalization,HollowoodLocalizationI,HollowoodLocalizationII}
noticed that the action functional
of the constrained instanton calculus for $\N=2$ super Yang-Mills 
has the form of cohomological field theory, to which 
powerful localization technique is applicable.
Based on this observation, Nekrasov and collaborators determined all the
instanton corrections and showed that they agree with the result obtained from the
Seiberg-Witten curve\cite{NekrasovCounting,NekrasovPartitions}.
He also proposed a simple extension of his formula  for
 the exact graviphoton-corrected  prepotential for the
five-dimensional (\five-dimensional) pure $U(N)$ super Yang-Mills theory.

We have also seen 
a great amount of
development on the side of topological strings in the last two years.
The advent of the so-called topological vertex enabled us to calculate
all-genus topological A-model amplitude for any local toric Calabi-Yau % manifold
using certain kind of Feynman-like rules.
According to \cite{BCOV,AGNT},
%Bershadsky \etal\cite{BCOV} and Antoniadis \etal\cite{AGNT},
 the topological amplitude should be equal to the
graviphoton-corrected prepotential of the physical theory
obtained by compactifying the type IIA string (or more precisely M-theory)
on the same Calabi-Yau.
Indeed, Iqbal and Kashani-Poor found \cite{IqbalKashaniPoorI,IqbalKashaniPoorII},
by using some physically-motivated simplifying assumptions,
that for  certain local toric Calabi-Yau $X^0_N$
the resulting all-genus amplitude exactly agreed with
the \five-dimensional\ prepotential  proposed by Nekrasov.
Later, Eguchi and Kanno \cite{EguchiKannoI,EguchiKannoII}
 showed in a mathematically rigorous way 
the equivalence of the two expressions. They also extended the results
to  cases with various matter contents.

However, certain  mysteries remain.
There are $N+1$  local
toric Calabi-Yau manifolds $X^m_N$ ($m=0,1,\ldots,N$)
which can be used to geometrically engineer the same \four-dimensional
$\N=2$ $U(N)$ super Yang-Mills theories.
As was shown in \cite{IqbalKashaniPoorII}, only for $m=0$ the topological 
amplitude agreed with the \five-dimensional\ prepotential proposed by Nekrasov.
What kind of physical mechanism is
operating behind the discrepancy for nonzero $m$?
We solve this puzzle in this short note.  Namely,
we exactly reproduce the all-genus closed string amplitudes
for $X^m_N$ with $m$ nonzero
in a gauge theory calculation \ala Nekrasov.
It is done by properly taking  into account
 the effects of  \five-dimensional\  Chern-Simons terms.

The rest of the paper is organized as follows. 
In section 2, we very briefly review the instanton counting by Nekrasov. 
In section 3, we review the relation between the \five-dimensional\ Chern-Simons
term and the triple intersection of the Calabi-Yau.
After these preparations, we analyze the effect of \five-dimensional\ Chern-Simons
terms to the instanton counting in section 4.
Finally, in section 5, we conclude by discussing some of the future directions.
In the appendix we discuss the relation between the partition functions on 
the $\Omega$ background
and the graviphoton-corrected prepotential.

\section{Brief Review of Nekrasov's Instanton Counting}
\label{review}
Let us recapitulate briefly the method of computation of the 
graviphoton-modified prepotential, as presented by Nekrasov.
We consider \five-dimensional\ pure Yang-Mills theory with eight supercharges.
We put it on the so-called $\Omega$ background with the metric
\begin{equation}
ds^2=(dx^5)^2+(dx^\mu+A^\mu dx^5)^2,
\end{equation}where $\Omega_{\mu\nu}=\partial_{[\mu}A_{\nu]}$ is
 constant anti-self-dual(ASD), and the circumference of the fifth direction  is $\beta$.
 We denote the eigenvalues of $\Omega_{\mu\nu}$ by $\pm\hbar$.

Firstly, we encode the vacuum expectation value of the adjoint scalar
as the Wilson line $\exp(\beta \diag (a_1,\ldots,a_N))$ around the fifth direction.
This is possible because \four-dimensional\ scalar field in the vector multiplet
consists of one real \five-dimensional\ scalar and the Wilson line around the circle.
From holomorphy, 
we have only to calculate the prepotential with Wilson line turned on,
with no vacuum expectation value for the   \five-dimensional\ scalar.

Secondly, we note that the partition function $Z$ of supersymmetric theories on
the $\Omega$ background 
and the graviphoton-corrected prepotential $\F_{\text{grav}}$ of the same theory 
generically satisfy the relation\begin{equation}
\F_{\text{grav}}=\hbar^2\log Z.\label{mysterious}
\end{equation}This relation was shown originally in \cite{NekrasovCounting}.
Another derivation using the Hilbert scheme of points is presented in the appendix.
$Z$ can be schematically expressed as\begin{equation}
Z=\tr (-)^F e^{-\beta H} e^{\beta \Omega_{\mu\nu}J^{\mu\nu}}
e^{\beta a_i J^i},
\end{equation}where $H$ is the Hamiltonian of the field theory,
$J^{\mu\nu}$ the generators of $SO(4)$, and
$J^i$ are the generators of global $U(N)$ gauge rotations.

Thirdly, it is argued that the calculation of the partition function
localizes onto the moduli space of instantons. Thus $Z$ is given by
\begin{equation}
Z=\sum_k q^{k} \tr (-)^F e^{-\beta H_k}
e^{-\beta \Omega_{\mu\nu}J^{\mu\nu}}e^{-\beta a_i J^i}\label{bar}
\end{equation} where $q=e^{2\pi i \beta \tau}$ counts the instanton number.
Now $H_k$ is the Hamiltonian of the supersymmetric
quantum mechanics on the ``framed'' $k$-instanton moduli space $\M_{N,k}$
of dimension $4Nk$.
From the index theorem, there are $4Nk$ real fermionic adjoint zero-modes 
around the $k$-instanton configuration.
Hence, the Hilbert space of the quantum mechanical system is the
space of sections of the spin bundle of the instanton moduli.
Thus, the trace in equation \eqref{bar} 
can be identified with the equivariant index of the $k$-instanton moduli.
We focus on the Cartan subgroup
$U(1)^{2+N}\subset SO(4)\times U(N)$ of the global symmetries.

We use the following fixed point theorem to calculate the equivariant index:\\
\textbf{Theorem}\quad Let $M$ be a spin manifold acted by an abelian
group $G$. Take an element $a$ from the Lie algebra of $G$ and let $g=e^{\beta a}$.
Assume the fixed points of $G$ are isolated.
Then\begin{equation}
\tr (-)^F g  = \sum_{
\substack{p:\text{fixed}\\ \text{ points of}\ G}}
\prod_{\substack{w:\text{eigenvalues }\\ \text{of $a$ on}\ TM_p}}
\frac{1}{2\sinh \frac\beta2 w }
\end{equation} where the left hand side is  traced over zero-modes
of the Dirac operators on $M$.

This theorem reduces the calculation of $Z$ to the study of fixed points
in the moduli space.  The fixed points are at the small instanton
singularities in $\M_{N,k}$. 
Hence we need to use the moduli of noncommutative  instantons
$\widetilde{\M_{N,k}}$ as the ultraviolet regularization.
The result turns out to be independent of the noncommutativity parameter.
Fixed points in $\widetilde{\M_{N,k}}$ was
 studied by Nekrasov \cite{NekrasovCounting}
and Nakajima and Yoshioka\cite{NakajimaYoshioka}.
They are labeled by $N$-tuples of Young tableaux $(Y_1,\ldots, Y_N)$.
%We identify a Young tableau $Y$ and the partition: 
%$Y=(y_1\ge y_2\ge y_3\ge\cdots)$.
We denote the number of boxes of the $i$-th row of the tableau $Y$ by $y_i$.
The action of $g$ on the tangent space at the fixed points can be studied
straightforwardly, and gives the celebrated formula of Nekrasov\begin{equation}
\exp(\hbar^{-2}\F_{\text{grav}})
=\sum_{Y_1,\ldots,Y_N}
\left(\frac{q}{4^{N}}\right)^{\sum \ell_{Y_i}}
\prod^N_{l,n=1}\prod^\infty_{i,j=1}
 \frac{\sinh \frac\beta2 (a_{ln}+\hbar(y_{l,j}-y_{n,i}+j-i))}
 {\sinh \frac\beta2 (a_{ln}+\hbar(y_{l,j}-y_{n,i}))}\label{Nek}
\end{equation}where 
$\ell_Y$ is the number  of boxes in $Y$,
$\hbar$ is the magnitude of $\Omega_{\mu\nu}$, and $a_{ij}$ denotes $a_i-a_j$.
Note that the result is independent under the constant shift $a_i\to a_i+c$.

To determine  the $g$ action on the tangent space, we need to use the 
Atiyah-Drinfeld-Hitchin-Manin (ADHM) construction of instantons.
Let $\X$ denote the vector space 
\begin{equation}
\X=
(V^*\otimes V) \oplus (V^*\otimes V)\oplus (W^*\otimes V)\oplus (V^*\otimes W)
\label{X}
\end{equation}where $V=\C^k$ and $W=\C^N$.
Define a $U(k)$ action on $\X$ by letting $V$ and $W$ 
transform in the fundamental and trivial representations respectively.
$\widetilde{\M_{N,k}}$
is the hyperk\"ahler quotient of $\X$ by the $U(k)$ action. 
Hence, a fixed point $p=gp$ in $\widetilde{\M_{N,k}}$ corresponds to an
ADHM datum $x_p\in X$ fixed up to the $U(k)$ action:\begin{equation} 
gx_p=\phi_p(g)x_p
\end{equation}where $g\in U(1)^{2+N}$ and $\phi_p(g)\in U(k)$.
An essential part of the calculation leading to the formula \eqref{Nek} is
the property that at the fixed point labeled by $(Y_1,Y_2,\ldots,Y_N)$, 
$\phi(g)$ has $k$ eigenvalues given by \begin{equation}
\exp(-\beta (a_l + \hbar (i-j)))\quad\text{for each box}\ (i,j)\in Y_l 
\end{equation}
where $l$ runs from $1,\ldots,N$.

\section{Triple intersection and the Chern-Simons terms}
In this section we recapitulate how the \five-dimensional\ theories obtained by
M-theory compactification on $X^m_N$ differ from each other.
Firstly let us recall the generic structure of \five-dimensional\ supersymmetric
$U(1)^n$ gauge theory with eight supercharges\cite{Seiberg5D}.
Classically, the theory is specified by the prepotential
$\F=c_{ijk} a_ia_ja_k+\tau_{ij}a_ia_j$
of degree up to three. It is because the
third derivative of $\F$ gives the coefficient of 
the Chern-Simons term\begin{equation}
\int c_{ijk}A_i\wedge F_j \wedge F_k.\label{d5CS}
\end{equation}
This will not be gauge-invariant unless the third derivative is constant.

Secondly let us see how the coefficients $c_{ijk}$ is determined from the geometric
data, when the theory is realized by a M-theory compactification.
In this setup, \five-dimensional\ vector fields $A_i$ come from
the three-form field $C^{(3)}$ of the eleven-dimensional supergravity reduced along
two-cycles $C_i$ in the Calabi-Yau, $A_i=\int_{C_i} C^{(3)}$.
The \five-dimensional\ Chern-Simons term \eqref{d5CS} comes directly from the
eleven dimensional Chern-Simons coupling \begin{equation}
\int C^{(3)}\wedge (dC)^{(4)}\wedge (dC)^{(4)}.
\end{equation}Thus we see that the coefficient $c_{ijk}$ is precisely the
triple intersection of (the Poincar\'e duals of) two-cycles $C_i$.

When the Calabi-Yau space develops an ADE singularity through the collapse
of two cycles, there appears enhanced non-abelian gauge symmetry corresponding to
the ADE type of  the singularity. 
In such cases, the Chern-Simons coupling \eqref{d5CS} should be likewise enhanced to the
non-abelian version $CS(A,F)$ which is defined through
 the descent equation\begin{equation}
d CS(A,F)= \tr (F\wedge F\wedge F)
\end{equation}where $F$ is the non-abelian field strength.
Moreover, Intriligator \etal \cite{IntriligatorMorrisonSeiberg}
studied the geometry of general Calabi-Yau manifolds which give rise to
\five-dimensional\ $U(N)$ theories and showed that the triple intersection
is determined up to the coefficient of this non-abelian Chern-Simons term.

Iqbal and Kashani-Poor studied
M-theory compactification on local toric Calabi-Yau manifolds $X^m_N$
\cite{IqbalKashaniPoorII}.
We collect here relevant facts on those manifolds. %without proof.
See \cite{IqbalKashaniPoorII,IntriligatorMorrisonSeiberg}
for more detailed accounts.
$X^m_N$ is a fibration of  $A_{N-1}$ singularity over the base $\CP^1$.
It contains a sequence of compact divisors \begin{equation}
S_1= \HF_{m+2-N},\quad S_2=\HF_{m+4-N},\quad \ldots,\quad
S_{N-1}=\HF_{m+N-2}
\end{equation}Here $\HF_n$ denotes a Hirzebruch surface.
The Hirzebruch surface $\HF_m$ is a $\CP^1$ fibration over the $\CP^1$
with the intersection pairing
\begin{equation}
B\cdot B=-m,\qquad
B\cdot F=1,\qquad
F\cdot F=0
\end{equation}where $B$ and $F$ denote the base and the fiber, respectively.
An $A_{N-1}$ singularity contains at the tip $N-1$ $\CP^1$'s
$C_1,\ldots,C_{N-1}$ with intersection pairing $C_i\cdot C_{i+1}=1$.
The divisor $S_i$ of $X^m_N$ is the fibration of $C_i$ over the base $\CP^1$.
The classical prepotential, or equivalently the triple intersection
 is given by the formula\begin{equation}
\F=\frac12\sum_{i,j}| a_i-a_j| ^3 + m\sum a_i^3
\end{equation}where we defined the basis $a_1,\ldots,a_N$
of special coordinates by \begin{equation}
\F=\sum_{i,j,k}(a_{i+1}+\cdots+a_N)(a_{j+1}+\cdots+a_N)
(a_{k+1}+\cdots+a_N)(S_i\cdot S_j\cdot S_k).
\end{equation} This is a natural choice since $S_i$ corresponds to
the simple root $a_i-a_{i+1}$.
From these expressions we see that the label $m$ of $X^m_N$ is exactly
proportional to the magnitude of the \five-dimensional\ Chern-Simons term.

\section{Effect of the Five-dimensional Chern-Simons term}
We saw in the last section that the M-theory compactifications on manifolds $X^m_N$
have different coefficients for the non-abelian \five-dimensional\ Chern-Simons terms
for different $m$.
Let us next see how the presence of \five-dimensional\ Chern-Simons terms
changes the derivation of the graviphoton-corrected prepotential
reviewed in section \ref{review}. 

\subsection{Calculation \ala Nekrasov}
Let us calculate the graviphoton-corrected prepotential of \five-dimensional\ 
$U(N)$ super Yang-Mills with $m$ units of non-abelian Chern-Simons term.
Firstly we put the theory on the $\Omega$ background and
encode the moduli of the theory
by introducing Wilson lines along the fifth direction.
The calculation is then
localized to that of supersymmetric quantum mechanics on the
ASD instanton moduli space.
The Lagrangian of the quantum mechanical system 
is obtained by substituting the gauge field in the \five-dimensional\ action
by the corresponding anti-self-dual configurations specified by the
trajectory in the moduli space.
The Yang-Mills action gives the kinetic term for the point particle moving on the ASD moduli,
and the Chern-Simons term gives a phase depending on the trajectory:\begin{equation}
e^{im\int CS(A,F) } = e^{ i m \int dx^i \A_i }
\end{equation} where $x^\mu(t)$ is the trajectory in the moduli space and 
the point particle is now coupled with an external vector potential $\A$.
Therefore, the exact partition function of the theory put on the $\Omega$ background is
\begin{equation}
Z_{\text{gauge}}=\sum_k q^{k} \tr (-)^F e^{-\beta H_k}
e^{-\beta \Omega_{\mu\nu}J^{\mu\nu}}e^{-\beta a_i J^i}.\label{boo}
\end{equation} Now $H_k$ is the Hamiltonian of the supersymmetric
quantum mechanics on $\M_{N,k}$ coupled to the external gauge potential $\A$.
The Hilbert space of the system is the space of sections of $S\otimes L$,
where $S$ is the spin bundle of $\M_{N,k}$ and
$L$ is the line bundle determined by $\A$.
Now we use the extended version of the
 fixed point theorem(see \textit{e.g.} \cite{Goodman}):\\
\textbf{Theorem} \quad
Let $E\to M$ be a vector bundle over a spin manifold with an
action by an abelian group $G$.
Let $a$ an element of the Lie algebra of $G$ and take $g=e^{\beta a}$.
Assume the fixed points of $G$ are isolated.
Then\begin{equation}
\tr (-)^F g 
= \sum_{p:\text{fixed points of}\ G}
(\tr g|_{E_p})
\prod_{\substack{w:\text{eigenvalues }\\ \text{of $a$  on}\ TM_p}}
\frac{1}{2\sinh \frac\beta2 w }
\end{equation}where the trace in the left hand side is taken over the zero-modes of
the Dirac operator on the spin bundle tensored by $E$.

In view of the theorem,
the study of $g$ action on $L|_p$  suffices to determine $Z$,
since we know the placement of fixed points and $g$ action on the tangent spaces already.
The line bundle $L$ has long been known to physicists.
It is the determinant line bundle $\Det \slash{D}$.
The determinant line bundle is defined on the space of connections $\A/\G$
and the fiber at a configuration $A$ is defined by
$(\det\Ker \slash{D}_A)^*\otimes \det\Ker \slash{D}_A^\dagger$ where
$D_A$ is the chiral Dirac operator coupled to the connection $A$
in the fundamental representation.
When the base space is restricted to the ASD moduli space, it can be simplified to
$(\det \Ker \slash{D}_A)^*$ because we know
 that there are no wrong chirality zero-modes.
Close relation between the determinant line bundle and the 
Chern-Simons terms has long been known since the work of Alvarez-Gaum\'e and Ginsparg
\cite{AlvarezGaumeGinsparg} on the geometric interpretation of
non-abelian anomalies. 

In reality we need to blow up the small instanton singularity in the ASD moduli
using spacetime non-commutativity. Hence we need the non-commutative
extension of all these relations among anomaly, index theorem and
the Chern-Simons terms.
Fortunately every detail we need has already been
worked out by various groups following the seminal work of
Seiberg and Witten on noncommutativity.
We refer the reader the works \cite{GraciaBondiaMartin,Bonora}
for noncommutative extension of
the relation of non-abelian anomalies,
Chern-Simons terms and the index theorem in six dimesions,
and the work \cite{NCzeromodes} for the study of the Dirac zero-modes
in the non-commutative instanton background.
We know from these works that there is no essential difference 
between commutative and non-commutative spacetime
with regard to the relation of anomaly and the Chern-Simons terms.

%\subsection{Calculation \ala Nekrasov, part II}
Now that we have clear understanding
on the nature and the structure of the line bundle
$L$, we can complete the calculation.
The fiber at $p$ is the highest exterior power of the kernel of the Dirac operator.
Thus, to determine the weight $w$, we have to determine the action of $g$
on the Dirac zero-modes.
%Fortunately, we know already the location of the fixed points
% on the instanton moduli space,
%and we know how to construct the Dirac zero-modes from the ADHM data.
As reviewed in section \ref{review},
the ADHM datum $x_p$
itself is not invariant under the action of $g$,
it maps $x_p$ to a datum equivalent under $U(k)$ 
transformation $\phi$:\begin{align}
gx_p=\phi(g)x_p.
\end{align}
Furthermore, we know from the analysis in \cite{NCzeromodes} that
$k$  Dirac zero-modes in the fundamental representation of $U(N)$
transforms as  a fundamental representation of $U(k)$.
One way of seeing this is to note that when
one reconstructs the ADHM datum from an ASD connection,
the $k$-dimensional vector space $V$ in equation \eqref{X} is none other than the space of
zero-modes of the Dirac equation.
These arguments show that the action of $g$ on the zero-modes can be traded by the 
action of $\phi(g)$. 
Hence $g|_{L_p}$ can be readily computed to give\begin{equation}
g|_{L_p}=\exp\left(-\beta\sum_k\sum_{(i,j)\in Y_k} (a_k+\epsilon(i-j))\right)
=\exp\left(-\beta\sum_k \left(\ell_{Y_k}a_k +\hbar\kappa_{Y_k}\right)\right)
\end{equation} where $(Y_1,\ldots,Y_N)$ labels the fixed points
and we defined \begin{equation}
\kappa_{Y}\equiv\sum_{(i,j)\in Y} (i-j)=\sum_i y_i(y_i+1-2i).
\end{equation}

Combining all these, we get the partition function for the \five-dimensional\ theory with
non-abelian Chern-Simons term:\begin{equation}
Z_{\text{gauge}}
=\sum_{Y_1,\ldots,Y_N}\left(\frac{q}{4^N}\right)^{\sum_i \ell_{Y_i}}
e^{-m\beta\sum (\ell_{Y_i}a_i+\hbar\kappa_{Y_i})}
\prod^N_{l,n=1}\prod^\infty_{i,j=1}
 \frac{\sinh \frac\beta2 (a_{ln}+\hbar(y_{l,j}-y_{n,i}+j-i))}
 {\sinh \frac\beta2 (a_{ln}+\hbar(y_{l,j}-y_{n,i}))}.\label{ala}
\end{equation}We defined $a_{ij}=a_i-a_j$ for brevity.
Note that the combined transformation\begin{equation}
a_i\to a_i+c,\qquad
2\pi i\tau \to 2\pi i\tau - mc\label{glsym}
\end{equation} does not change the result as it should be. 
%One can assume $\sum a_i=0$ using this global symmetry.

\subsection{Comparison with the topological A-model amplitudes}
Let us compare what we have obtained \ala Nekrasov against the topological A-model
amplitudes for local toric Calabi-Yau manifolds $X^m_N$. 
Combining the equations (67,68,69,78)
 in the article by Iqbal and Kashani-Poor\cite{IqbalKashaniPoorII}
 and changing their notation to ours, 
 the amplitude  is\begin{multline}
Z_{\text{top.}}=\sum_{Y_1,\ldots,Y_N}
%M^{(m)}(q,Y_i)
2^{-2N\sum \ell_{Y_i}}
(-)^{(N+m)\sum \ell_{Y_i}} q^{\frac12\sum_{i=1}^{N}(N+m-2i)\kappa_i}
%\prod^N_{i=1} Q^{l_{Y_i}}_{b_i}
Q_B^{\sum \ell_i}\times\\
\prod_{i=1}^{\lfloor \frac{N+m-1}2\rfloor} Q_i^{(N+m-2i)(\ell_1+\cdots+\ell_i)}
\prod_{i=\lfloor \frac{N+m+1}2\rfloor}^{N-1}
Q_i^{(2i-m-N)(\ell_{i+1}+\cdots+\ell_N)}
\prod^{N-1}_{i=1}Q_{b_i}^{-(N-i)(\ell_1+\cdots+\ell_i)-i(\ell_{i+1}+\cdots+\ell_N)}\\
\times e^{-\frac12\beta\hbar\sum^N_{i=1}(N-2i)\kappa_i}
\prod^N_{l,n=1}\prod^\infty_{i,j=1}
 \frac{\sinh \frac\beta2 (a_{ln}+\hbar(y_{l,j}-y_{n,i}+j-i))}
 {\sinh \frac\beta2 (a_{ln}+\hbar(y_{l,j}-y_{n,i}))}
\end{multline}where $Q_B$ and $Q_i=e^{-\beta(a_i-a_{i+1})}$ are
respectively the exponential of the K\"ahler parameters
of the base divisor and the divisors $S_i$.
Define $a_N$ by \begin{equation}
e^{-\beta a_N}= -\left(Q_B
\prod_{i=1}^{\lfloor \frac{N+m-1}2\rfloor} Q_i^{-i}
\prod_{i=\lfloor \frac{N+m+1}2\rfloor}^{N-1} Q_i^{-(N-i+m)}
\right)^{\frac1m}.
\end{equation}
Then, after reshuffling the various factors with some effort, 
one finds that \begin{equation}
Z_{\text{top.}}=
\sum_{Y_1,\ldots,Y_N}\frac{1}{(-4)^{N{\sum \ell_{Y_i}}}}
e^{-m\beta\sum (\ell_{Y_i}a_i+\hbar\kappa_{Y_i})}
\prod^N_{l,n=1}\prod^\infty_{i,j=1}
 \frac{\sinh \frac\beta2 (a_{ln}+\hbar(y_{l,j}-y_{n,i}+j-i))}
 {\sinh \frac\beta2 (a_{ln}+\hbar(y_{l,j}-y_{n,i}))}.
\end{equation}
Using the global symmetry \eqref{glsym},
we see that this precisely agrees with the calculation of the
gauge theory side \ala Nekrasov, equation \eqref{ala}.

\section{Conclusion}
In this short note,
we extended the \five-dimensional\ graviphoton-corrected prepotential
proposed by Nekrasov to include the effect of \five-dimensional\  
non-abelian Chern-Simons term.
We saw  that the introduction of 
\five-dimensional\ Chern-Simons terms  results in 
twisting of the spin bundle  on the instanton moduli
by the determinant line bundle.
We obtained using the fixed point theorem
the partition function of the
\five-dimensional\ $U(N)$ super Yang-Mills theory with
\five-dimensional\ Chern-Simons term
on the $\Omega$ background.
Moreover we  saw that the result 
precisely reproduced the topological string amplitude
originally obtained by
Iqbal and Kashani-Poor and mathematically proved by Eguchi and Kanno.
This agreement is as it should be, because the partition function of the 
topological A-model on a Calabi-Yau and that of M-theory
compactified on the same Calabi-Yau times the $\Omega$ background should be equal.

It will be worthwhile to generalize the calculation \ala Nekrasov to
general \five-dimensional\ theories obtained from M-theory compactification on
Calabi-Yau manifolds.  It will be extremely interesting if we can generically prove
the equality of the gauge theory partition function with the topological string
partition function by extending Nekrasov's instanton counting.
As the topological string amplitudes can be calculated from
the topological vertex, it will be tempting to suggest the existence
of some kind of `gauge vertex' which upon gluing yields general gauge theory
partition functions.

\paragraph{Acknowledgment}
The author thanks to K. Sakai, R. Nobuyama, Prof.\ T. Eguchi and Prof.\ H. Kanno
for very stimulating discussion on the subject.

\section*{Appendix: Derivation of the relation \eqref{mysterious}}
In this appendix we give yet another derivation of the relation \eqref{mysterious}.
Firstly recall the argument presented by Gopakumar and Vafa.
They showed in
\cite{GopakumarVafaI,GopakumarVafaII} that an BPS multiplet with the left spin content
\begin{equation}
I_r=\left((\frac12)\oplus 2(0)\right)^{r+1}
\end{equation} and with central charge $a$ contributes to
the graviphoton-corrected prepotential by an amount
\begin{equation}
\hbar^{-2}\F_r(a)=\sum_{k>0}\frac{1}{k}(2\sinh\frac{k \hbar}{2})^{2r-2} e^{-ka}
\end{equation}where $\hbar$ is the magnitude of the field strength
of the graviphoton.
This equation can be proved using the Fock-Schwinger proper time method.
Another convenient basis of the left spin content is \begin{equation}
C_j=\left((\frac12)\oplus 2(0)\right)\otimes( \text{a state with}\ J^3_L=j)
\end{equation}In this basis, the contribution to the prepotential becomes\begin{align}
\hbar^{-2}\F_j(a)&=
\sum_{k>0}\frac{1}{k}\frac1{(2\sinh k\hbar /2)^2} e^{-k(a+2j\hbar )}\\
&=\sum_{n>0}\log\left(1-e^{-(a+2j\hbar +n\hbar )}\right).
\end{align}
The prepotential is a quantity protected by supersymmetry and receives contributions
only from states annihilated by half of the supersymmetry, \textit{i.e.}\ BPS states.
Thus the exact prepotential of the low energy theory is given by\begin{equation}
\hbar^{-2}\F=\sum_{i,r} N_{i,r}
\sum_{k>0}\frac{1}{k}(2\sinh\frac{k \hbar}{2})^{2r-2} e^{-ka_i}
\label{grav-prep}
\end{equation}where $N_{i,r}$ is the number of multiplets with central charge $a_i$
and spin content $I_r$.  $N_{i,r}$ is called the Gopakumar-Vafa invariants
of the theory.

Next consider the partition function of the theory on the $\Omega$ background.
Let us canonically quantize the theory,
considering the fifth direction $dx^5$ as the time direction.
Then, in the Hamiltonian formalism, the partition function can be schematically written as
\begin{equation}
Z=\tr (-)^F e^{-\beta H} \exp(i\beta \Omega_{\mu\nu} J^{\mu\nu}).
\end{equation}where $H$ is the total Hamiltonian of the field theory.
This is none other than the equivariant index of the system.
$\exp(i\beta \Omega_{\mu\nu} J^{\mu\nu})$ commutes with 
half of the supersymmetry when the curvature $\Omega_{\mu\nu}$ is self-dual.
Thus, the partition function $Z$ receives contributions only from the states
annihilated by those supersymmetry. These states are precisely what contributed
to the prepotential of the theory put on the graviphoton background.
These consideration reveals us that the partition function can be written as an
infinite product of the form\begin{equation}
Z=\prod_{r,i} Z_r(a_i)^{N_{i,r}}
\end{equation}where $N_{i,r}$ is the same Gopakumar-Vafa invariants we discussed above.
Hence, we need only to show\begin{equation}
Z_r(a)=\exp(\hbar^{-2}\F_r(a))\label{basic}
\end{equation}in order to show the relation \eqref{mysterious}.

Let us show the relation \eqref{basic} for the case $r=0$.
The extension to other cases should be immediate.
Hence we are going to calculate the partition function of a free hypermultiplet
on the $\Omega$ background.
In a first-quantized framework, the system is thought of as a collection of particles and 
anti-particles.  The calculation of the partition function is localized by the 
supersymmetry to the configuration space of BPS states.
A BPS configuration is a collection of particles only, since an anti-particle
respects the other half of the supersymmetry and particle-antiparticle pair breaks
all of the supersymmetry.
As the particles are indistinguishable from each other, the configuration space of $k$ particles
is \begin{equation}
S^k\C^2\equiv (\C^2)^k/\S_k.
\end{equation}
Hence, the partition function should be\begin{equation}
Z=\sum e^{-ka}\Ind_g S^k\C^2.
\end{equation}where $\Ind_g M$ denotes the equivariant index,
i.e. the trace of $g$ over the space of harmonic spinors of $M$.
However, the space $S^k\C^2$ is singular and reliable calculation of the 
equivariant index is difficult.
It is known that there is a good resolution $(\C^2)^{[k]}$
of the singularities of $S^k\C^2$,
called the Hilbert scheme of $k$-points on $\C^2$. 
Furthermore, it is known to be identical to $\widetilde{M_{1,k}}$,
the moduli of non-commutative $U(1)$ instantons.
Hence, the fixed points in $(\C^2)^{[k]}$ is labeled by a Young tableau $Y$
and the result is \begin{equation}
Z=\sum_Y e^{-\ell_Y a}\prod_{s\in Y}\left(\frac{1}{\sinh\frac
\beta2\hbar (l(s)+a(s)+1)}\right)^2 \label{atype}
\end{equation}where
we defined for a Young tableau its arm length and leg length by
\begin{align}
a_Y(s)&=y_i-j,&
l_Y(s)&=y^D_j-i.
\end{align} for a box $s=(i,j)\in Y$. $Y^D$ denotes the transpose of the
Young tableau $Y$.
 We refer the reader to the lecture notes by H. Nakajima\cite{NakajimaLecture}
for a detailed derivation.
The expression \eqref{atype} can be simplified using the 
free fermion technique \cite{EguchiKannoII} to give\begin{align}
Z&=\prod_{n\ge 1}\left(\frac1{1-e^{-a}e^{-\hbar n}}\right)^n
=\exp(\hbar^{-2}\F_{r=0}(a)).
\end{align} %This is what we wanted to derive.


\begin{thebibliography}{99}

    \bibitem{FlumeLocalization}

R.~Flume, R.~Poghossian and H.~Storch,
 ``The Seiberg-Witten prepotential and the Euler class of the reduced  moduli
space of instantons,'' {\slshape 
Mod.\ Phys.\ Lett.\ A }{\bf 17}, 327 (2002)
[arXiv:hep-th/0112211].
%%CITATION = HEP-TH 0112211;%%

    \bibitem{HollowoodLocalizationI}

T.~J.~Hollowood,
 ``Calculating the prepotential by localization on the moduli space of
instantons,'' {\slshape 
JHEP }{\bf 0203}, 038 (2002)
[arXiv:hep-th/0201075].
%%CITATION = HEP-TH 0201075;%%

    \bibitem{HollowoodLocalizationII}

T.~J.~Hollowood,
``Testing Seiberg-Witten theory to all orders in the instanton expansion,'' {\slshape 
Nucl.\ Phys.\ B }{\bf 639}, 66 (2002)
[arXiv:hep-th/0202197].
%%CITATION = HEP-TH 0202197;%%

    \bibitem{NekrasovCounting}

N.~A.~Nekrasov,
``Seiberg-Witten prepotential from instanton counting,''
arXiv:hep-th/0206161.
%%CITATION = HEP-TH 0206161;%%

    \bibitem{NekrasovPartitions}

N.~Nekrasov and A.~Okounkov,
``Seiberg-Witten theory and random partitions,''
arXiv:hep-th/0306238.
%%CITATION = HEP-TH 0306238;%%

    \bibitem{BCOV}

M.~Bershadsky, S.~Cecotti, H.~Ooguri and C.~Vafa,
 ``Kodaira-Spencer theory of gravity and exact results for quantum string
amplitudes,'' {\slshape 
Commun.\ Math.\ Phys.\  }{\bf 165}, 311 (1994)
[arXiv:hep-th/9309140].
%%CITATION = HEP-TH 9309140;%%

    \bibitem{AGNT}

I.~Antoniadis, E.~Gava, K.~S.~Narain and T.~R.~Taylor,
``Topological amplitudes in string theory,'' {\slshape 
Nucl.\ Phys.\ B }{\bf 413}, 162 (1994)
[arXiv:hep-th/9307158].
%%CITATION = HEP-TH 9307158;%%

    \bibitem{IqbalKashaniPoorI}

A.~Iqbal and A.~K.~Kashani-Poor,
``Instanton counting and Chern-Simons theory,''
arXiv:hep-th/0212279.
%%CITATION = HEP-TH 0212279;%%

    \bibitem{IqbalKashaniPoorII}

A.~Iqbal and A.~K.~Kashani-Poor,
``SU(N) geometries and topological string amplitudes,''
arXiv:hep-th/0306032.
%%CITATION = HEP-TH 0306032;%%

    \bibitem{EguchiKannoI}

T.~Eguchi and H.~Kanno,
``Topological strings and Nekrasov's formulas,'' {\slshape 
JHEP }{\bf 0312}, 006 (2003)
[arXiv:hep-th/0310235].
%%CITATION = HEP-TH 0310235;%%

    \bibitem{EguchiKannoII}

T.~Eguchi and H.~Kanno,
 ``Geometric transitions, Chern-Simons gauge theory and Veneziano type
amplitudes,''
arXiv:hep-th/0312234.
%%CITATION = HEP-TH 0312234;%%

    \bibitem{NakajimaYoshioka}

H.~Nakajima and K.~Yoshioka,
``Instanton counting on blowup. I,''
arXiv:math.ag/0306198.
%%CITATION = MATH-AG 0306198;%%

    \bibitem{Seiberg5D}

N.~Seiberg,
 ``Five dimensional SUSY field theories, non-trivial fixed points and  string
dynamics,'' {\slshape 
Phys.\ Lett.\ B }{\bf 388}, 753 (1996)
[arXiv:hep-th/9608111].
%%CITATION = HEP-TH 9608111;%%

    \bibitem{IntriligatorMorrisonSeiberg}

K.~A.~Intriligator, D.~R.~Morrison and N.~Seiberg,
 ``Five-dimensional supersymmetric gauge theories and degenerations of
Calabi-Yau spaces,'' {\slshape 
Nucl.\ Phys.\ B }{\bf 497}, 56 (1997)
[arXiv:hep-th/9702198].
%%CITATION = HEP-TH 9702198;%%

    \bibitem{Goodman}

M.~W.~Goodman,
``Proof Of Character Valued Index Theorems,'' {\slshape 
Commun.\ Math.\ Phys.\  }{\bf 107}, 391 (1986).
%%CITATION = CMPHA,107,391;%%

    \bibitem{AlvarezGaumeGinsparg}

L.~Alvarez-Gaum\'e and P.~Ginsparg,
``The Topological Meaning Of Nonabelian Anomalies,'' {\slshape 
Nucl.\ Phys.\ B }{\bf 243}, 449 (1984).
%%CITATION = NUPHA,B243,449;%%

    \bibitem{GraciaBondiaMartin}

J.~M.~Gracia-Bondia and C.~P.~Martin,
``Chiral gauge anomalies on noncommutative $\mathbb{R}^4$,'' {\slshape 
Phys.\ Lett.\ B }{\bf 479}, 321 (2000)
[arXiv:hep-th/0002171].
%%CITATION = HEP-TH 0002171;%%

    \bibitem{Bonora}

L.~Bonora, M.~Schnabl and A.~Tomasiello,
``A note on consistent anomalies in noncommutative YM theories,'' {\slshape 
Phys.\ Lett.\ B }{\bf 485}, 311 (2000)
[arXiv:hep-th/0002210].
%%CITATION = HEP-TH 0002210;%%

    \bibitem{NCzeromodes}

K.~Y.~Kim, B.~H.~Lee and H.~S.~Yang,
``Zero-modes and Atiyah-Singer index in noncommutative instantons,'' {\slshape 
Phys.\ Rev.\ D }{\bf 66}, 025034 (2002)
[arXiv:hep-th/0205010].
%%CITATION = HEP-TH 0205010;%%

    \bibitem{GopakumarVafaI}

R.~Gopakumar and C.~Vafa,
``M-theory and topological strings. I,''
arXiv:hep-th/9809187.
%%CITATION = HEP-TH 9809187;%%

    \bibitem{GopakumarVafaII}

R.~Gopakumar and C.~Vafa,
``M-theory and topological strings. II,''
arXiv:hep-th/9812127.
%%CITATION = HEP-TH 9812127;%%

    \bibitem{NakajimaLecture}

H. Nakajima, ``Lectures on Hilbert Schemes of Points on Surfaces'', 1999, AMS

\end{thebibliography}
\end{document}